\documentclass[prd,superscriptaddress,nofootinbib,showpacs,showkeys,preprint]{revtex4}
\usepackage{graphicx}
\usepackage[usenames,dvipsnames]{color}
\usepackage{amsmath,amssymb}
\usepackage{mathtools}
\usepackage[colorlinks]{hyperref}
\usepackage{float}

\newcommand{\be}{\begin{eqnarray}}                                             
\newcommand{\ee}{\end{eqnarray}}

\usepackage{epsfig}

\newcommand{\noplus}{}

\newcommand{\tmem}[1]{{\em #1\/}}
\newcommand{\tmop}[1]{\ensuremath{\operatorname{#1}}}

\begin{document}
\title{Absence of  the Gribov ambiguity in a quadratic gauge}
\author{Haresh Raval}
\email{haresh@phy.iitb.ac.in}
\affiliation{Department of Physics, Indian Institute of Technology, 
Bombay, Mumbai - 400076, India}
\begin{abstract}
 The Gribov ambiguity  exists in various gauges except algebraic gauges. However, algebraic gauges are not Lorentz invariant, which is their fundamental flaw. In addition, they are not generally compatible with the boundary conditions on the gauge fields, which are needed to compactify the space i.e., the ambiguity continues to exist on a compact manifold. Here we discuss a quadratic gauge fixing, which is Lorentz invariant.  We consider an example of a spherically symmetric gauge field configuration in which we prove that this Lorentz invariant gauge removes the ambiguity on a compact manifold $\mathbb{S}^3$, when a proper boundary condition on the gauge configuration is taken into account. Thus,  providing one  example where the ambiguity is absent on a compact manifold in the algebraic gauge. We also show that the \tmem{BRST} invariance is  preserved in this gauge.
\end{abstract}
\maketitle
\section{Introduction}
Defining the path integral in gauge theories has a major issue of infinite redundant functional integrations. The fact that the Yang-Mills action 
is invariant under the gauge transformation
is the cause of the issue. The issue is addressed by invoking a gauge condition such as the Landau gauge $\partial^\mu A_\mu=f$. However, it is shown in ref.~\cite{1} that even after the Landau gauge fixing, there  still exist  equivalent configurations, which contribute to the measure of the path integral. This implies that the Landau gauge  does not uniquely choose a configuration, the problem known as the Gribov ambiguity. We need only inequivalent configurations in the measure in order to properly quantize the theory.  The  inequivalent configurations can be extracted out by restricting the space of integration to the fundamental modular region $C^0$, where the faddeev-poppov operator has positive eigenvalues~\cite{1}. However, the region $C^0$ still contains  Gribov copies~\cite{1}.  The restriction on the space of integration is achieved by adding suitable terms to the effective action $S_{\tmop{eff}}$ resulting from the Landau gauge fixing~\cite{2,3}. This modified action is known as Gribov-Zwanziger action. The GZ action is not \tmem{BRST} invariant~\cite{4}. So, in an attempt to eliminate the Gribov copies, we lose the \tmem{BRST} invariance of the theory.  The same ambiguity is shown to exist in all the covariant gauges~\cite{5}. 

An essential reason why some gauges have the ambiguity is the differential operator involved in the gauge. Algebraic gauges are ambiguity free since they do not have a differential operator, but they have one disadvantage. In general, they violate the Lorentz invariance, which is a basic requirement for any theory. Whereas, the gauge under consideration in this paper is Lorentz invariant. It also turns out that, the theory is \tmem{BRST} invariant. Alternative formulations addressing the Gribov ambiguity  are suggested in  ref.~\cite{6,7}.  The former ref. particularly is an approach using Lorentz invariant algebraic gauge conditions. 

The contents of this paper are arranged as follows: in the next section, we discuss a particular quadratic gauge and its consequences at infra-red scale. In sec.~\ref{s4}, we examine a case of the spherically symmetric gauge  configuration. We prove that when a proper boundary condition on the gauge  configuration at $\infty$ is taken into account, the  quadratic gauge uniquely chooses the configuration on a compact manifold $\mathbb{S}^3$.  

\section{a quadratic gauge and Effective Lagrangian}\label{sec1}
There have been studies using quadratic gauges  in several contexts. A few of the references are~\cite{8,9,10,11,12,13}.  Here we consider a particular quadratic gauge introduced in the ref.~\cite{14} in the context of non-perturbative phenomena in QCD. 
\begin{align} \label{eq: 0 }
 H^a [ A^{\mu} ( x) ] =
A^a_{\mu} ( x) A^{\mu a} ( x) = f^a ( x) ; \  \text{  for each $a$ }
\end{align}
where $f^a(x)$ is an arbitrary function of $x$. This gauge condition results in  the effective Lagrangian of the form \cite{14}
 \begin{eqnarray} \label{eq: 2 }
  \mathcal{L}_{\tmop{eff}} &=& \mathcal{L}_{\tmop{YM}}+ \mathcal{L}_{\tmop{GF}}+ \mathcal{L}_{ghost}\nonumber\\
 &=& - \frac{1}{4} F^a_{\mu \nu} F^{\mu \nu a}
\noplus - \frac{1}{2 \zeta}  ( A^a_{\mu} A^{\mu a})^2 - \overline{c^a}
A^{\mu a} ( D_{\mu} c)^a 
\end{eqnarray}
where the first term is Yang-Mills Lagrangian with 
$ F^a_{\mu \nu}(x) = \partial_{\mu}A^a_{\nu}(x)- \partial_{\nu}A^a_{\mu}(x)-g f^{abc} A^b_{\mu}(x)A^c_{\nu}(x)$, second and third terms are gauge fixing and ghost Lagrangian respectively and $( D_{\mu} c)^a = \partial_\mu c^a - g f^{a b c}A_\mu^b c^c$. In terms of auxiliary fields $F^a$,  the effective Lagrangian can be rewritten as
\be \label{eq: 9 }
\mathcal{L}_{\tmop{eff}} &=& \mathcal{L}_{\tmop{YM}} + \frac{\zeta}{2}F^{a 2} + F^a\ A^a_{\mu} A^{\mu a} - \overline{c^a}A^{\mu a} ( D_{\mu} c)^a 
\ee 

The ghost Lagrangian  contains a term $gf^{a b c} \overline{c^a}c^c A^{\mu a} A_\mu^b$. For each ghost bilinear $\overline{c^a}c^c$, one can introduce an auxiliary field $\sigma$ through a unity in the path integral as shown in~\cite{14}. The ghost $c^3$ can be given a propagator by an additional gauge fixing. Then,  auxiliary fields can be given the effective potential, which has nontrivial minima, by a Coleman-Weinberg mechanism
in which one-loop  diagrams give  the leading quantum
correction.  In the present case, one-loop $c^3$ diagrams give the leading contribution.  The vacuum of ghost bilinears $\langle\overline{c^a}c^c\rangle$  can be shown to correspond nontrivial minima of auxiliary fields~\cite{14}.  Thus, with an assumption that ghost bilinears   under go condensation as described, the term  $gf^{a b c} \overline{c^a}c^c A^{\mu^a} A_\mu^b$ can be  seen to provide the mass matrix for gluons. The mass matrix has $N(N-1)$ non-zero eigenvalues only and thus has nullity $N-1$~\cite{14}. The  non-zero eigenvalues correspond to massive off-diagonal gluons and nullity correspond to massless diagonal gluons. The massive off-diagonal gluons are presumed to provide an evidence of Abelian dominance. Thus  Abelian dominance, which itself is an indication to the confinement, is easily evident in this gauge. Moreover, the  off-diagonal gluon after getting mass acquires the propagator of the form
\be\label{89}
(\mathcal{O}_{\text{ofd}}^{- 1})_{\mu \nu}^{    a  b}  ( p) = -
\frac{i \ \delta^{a b}}{p^2- M_{gluon}^2}  \left( \eta_{\mu \nu} - \frac{p_{\mu} p_{\nu}}{M_{gluon}^2} \right)
\ee
Since a mass term for the off-diagonal gluon is purely imaginary~\cite{14}, the propagator has no poles on a real $p^2$ axis, which is a sufficient condition for the confinement~\cite{15}. 
Thus, the two strong signatures of the confinement: 1. Abelian dominance and 2. A pole of the off-diagonal gluon  propagator is on imaginary $p^2$ axis become visible as a result of employment of the gauge. We now turn to the example.

\section{Spherically symmetric gauge potential and the quadratic Gauge} \label{s4}
Here we demonstrate that  the quadratic gauge uniquely picks up a spherically symmetric  configuration on a compact manifold  $\mathbb{S}^{3}$, when a proper boundary condition on the field is required to be satisfied. Compactification of a
 euclidean space  $\mathbb{R}^N$ to a compact manifold $\mathbb{S}^{N}$ is achieved by the condition $U(\infty)=I$~\cite{5}.  Since the space in this example is $\mathbb{R}^3$, the condition would compactify it to $\mathbb{S}^{3}$.  We  begin by adopting a  parameterization for a vector potential shown in ref.~\cite{1} 
 \begin{equation}
 A_i=f_1(r)  \frac{\partial\hat{n}}{\partial x_i} +f_2(r)\hat{n} \frac{\partial\hat{n}}{\partial x_i} +f_3(r)\hat{n}n_i ,  \hspace{1 cm}   i=1,2,3
 \end{equation}
Where  $n_i= \frac{x_i}{r}$,  \  \  $r=\sqrt{\Sigma x_i^2}$,  \  \
$\hat{n}=in_j \sigma_j$ \  \  $\sigma_j$  are Pauli matrices , $\hat{n}^2 = -1$. For simplicity we choose $A_0=0$.
Now, the spherically symmetric operator is given by
\begin{equation}
 U=exp⁡\Big(\frac{\alpha(r)}{2}  \hat{n}\Big )=\cos⁡\Big(\frac{\alpha(r)}{2}\Big)+ \hat{n} \sin⁡\Big(\frac{\alpha(r)}{2}\Big)
\end{equation}
Therefore, the compactification condition $U(\infty)=I$ implies $\alpha(\infty)= 4 \pi n; \ n \ \text{is an integer}$.  
The gauge transformation $A_\mu  \longrightarrow \tilde{A}_\mu = UA_\mu U^{-1}+i(\partial_\mu U)U^{-1}$  results in transformations of $f_1, f_2$ and $f_3$ as follows\\
\begin{equation}\label{tra}
\begin{split}
\tilde{f_1}&=f_1  \cos⁡\alpha+(f_2+\frac{1}{2})  \sin⁡\alpha\\
\tilde{f_2}+\frac{1}{2}&=-f_1  \sin⁡\alpha+(f_2+\frac{1}{2})  \cos⁡\alpha\\
\tilde{f_3}&=f_3  ̇+\frac{1}{2} \dot{\alpha} \  
\end{split}
\end{equation}
where overdot indicates differentiation with respect to $r$. Now, $a\ th $ component of $A_i$ can be derived using following formula
\begin{align}\label{11}
 A_i^a &= \frac{1}{2}Tr(A_i\sigma_a) \nonumber\\
 &=\frac{1}{2}Tr \left(f_1(r)  \frac{\partial\hat{n}}{\partial x_i}\sigma_a +f_2(r)\hat{n} \frac{\partial\hat{n}}{\partial x_i}\sigma_a +f_3(r)\hat{n}n_i\sigma_a\right)
\end{align}
To evaluate Eq.~\eqref{11}, we need to evaluate following entities
\begin{align}\label{12}
  Tr\Big(\frac{\partial\hat{n}}{\partial x_i}\sigma_a\Big)&=i\frac{\partial n_j}{\partial x_i}Tr(\sigma_j \sigma_a)\nonumber \\
 &= i\frac{\partial n_j}{\partial x_i}Tr(\delta_{j a} +i \epsilon_{j a k}\sigma_k)\nonumber \\
 &=2i\frac{\partial n_a}{\partial x_i}
 \end{align}
  \begin{align}\label{13}
  Tr\Big( \hat{n} \frac{\partial\hat{n}}{\partial x_i}\sigma_a\Big)&=-Tr(n_q  \frac{\partial n_j}{\partial x_i} \sigma_q \sigma_j \sigma_a)\nonumber \\
  &= -n_q  \frac{\partial n_j}{\partial x_i} Tr\Big(i\epsilon_{j a k}(\delta_{q k}+i\epsilon_{q k l}\sigma_l)\Big)\nonumber \\
  &= -2i n_q  \frac{\partial n_j}{\partial x_i} \epsilon_{j a q}
\end{align}
\begin{align}\label{14}
  Tr (\hat{n}n_i\sigma_a) = 2in_in_a
\end{align}
Using Eq.s~\eqref{12}, \eqref{13}, \eqref{14} we find
\begin{subequations}\label{18}
\begin{align}
A_1^1 &=i[f_1 (\frac{1}{r}-\frac{x_1^2}{r^3}) + f_3\frac{x_1^2}{r^2}]\\
A_2^1&=i[-f_1\frac{x_1x_2}{r^3}+f_2\frac{x_3}{r^2}+f_3\frac{x_1x_2}{r^2}]\\
A_3^1&=i[-f_1\frac{x_1x_3}{r^3}-f_2\frac{x_2}{r^2}+f_3\frac{x_1x_3}{r^2}]
\end{align}
\end{subequations}
\begin{subequations}\label{19}
\begin{align}
A_1^2 &=i[-f_1\frac{x_1x_2}{r^3}-f_2\frac{x_3}{r^2}+f_3\frac{x_1x_2}{r^2}]\\
A_2^2&=i[f_1 (\frac{1}{r}-\frac{x_2^2}{r^3}) + f_3\frac{x_2^2}{r^2}]\\
A_3^2&=i[-f_1\frac{x_2x_3}{r^3}+f_2\frac{x_1}{r^2}+f_3\frac{x_2x_3}{r^2}]
\end{align}
\end{subequations}
\begin{subequations}\label{20}
\begin{align}
A_1^3 &=i[-f_1\frac{x_1x_3}{r^3}+f_2\frac{x_2}{r^2}+f_3\frac{x_1x_3}{r^2}]\\
A_2^3&=i[-f_1\frac{x_2x_3}{r^3}-f_2\frac{x_1}{r^2}+f_3\frac{x_2x_3}{r^2}]\\
A_3^3&=i[f_1 (\frac{1}{r}-\frac{x_3^2}{r^3}) + f_3\frac{x_3^2}{r^2}]
\end{align}
\end{subequations}
We now impose a boundary condition on $A_k^j$ s. We require that 
\begin{equation} \label{bou}
A_k^j \rightarrow 0  \ \ \text{as} \ \  \frac{1}{r}, \ \ \text{as} \  \ r\rightarrow \infty
\end{equation}
From Eq.s~\eqref{18}, \eqref{19}, \eqref{20}, it is clear that  this condition is achievable and the general boundary condition on $ f_1, f_2  \ \text{and} \  f_3$ can be easily interpreted, which is as following
 \begin{equation}\label{23}
 f_1, f_2 \rightarrow const. \ \ \text{as} \ \ r \rightarrow \infty  \text{ and}   \  f_3 \rightarrow 0 \ \ \text{as fast as}  \  \ \frac{1}{r} \ \ \text{as} \ \ r \rightarrow \infty
\end{equation} 
Here we make one note. We want to address the ambiguity on  $\mathbb{S}^{3}$, therefore a boundary condition on $f_3$ needs to be little stronger ($\text{faster than} \ \ \frac{1}{r} \ \ \text{as} \ \ r \rightarrow \infty$)  because of  the eq. for copies~\eqref{22} that we shall come across later in the section. Hence, we consider a stronger condition on $f_3$ only.
We will use these boundary conditions to prove our claim. We first
  evaluate a condition
\be
 A_i^a A^{i a}&=& A_1^aA^{1 a}+ A_2^aA^{2 a} +A_3^aA^{3 a};  \  \ \text{for each} \  a \nonumber
 \ee
 For example taking $ a = 1 $,  the gauge above takes the form
 \be
A_i^1A^{i 1}&=& A_1^1A^{1 1}+ A_2^1A^{2 1} +A_3^1A^{3 1} \nonumber\\
&=&(A_1^1)^2+(A_2^1)^2+(A_3^1)^2\nonumber\\
&=&-\Big[\frac{f_1^2}{r^2}(1-\frac{x_1^2}{r^2})+\frac{f_2^2}{r^2}(1-\frac{x_1^2}{r^2})+f_3^2\frac{x_1^2}{r^2}\Big]\nonumber
\shortintertext{In spherical polar coordinates, the condition can be written as }
&=&-\frac{1}{r^2}(f_1^2+f_2^2) + \sin^2\theta \cos^2\phi \Big(\frac{1}{r^2}(f_1^2+f_2^2)-f_3^2\Big)
\shortintertext{Hence, }\nonumber
\tilde{A}_{i}^1 \tilde{A}^{i 1}&=& -\frac{1}{r^2}(\tilde{f}_1^2+\tilde{f}_2^2) + \sin^2\theta \cos^2\phi \Big(\frac{1}{r^2}(\tilde{f}_1^2+\tilde{f}_2^2)- \tilde{f}_3^2\Big)
\ee 
The gauge equivalence $\tilde{A}_{i}^1 \tilde{A}^{i 1}=A_i^1A^{i 1}$ implies the following
\be\label{15}
 \frac{1}{r^2}[(\tilde{f}_1^2+\tilde{f}_2^2)-(f_1^2+f_2^2)]+\sin^2\theta \cos^2\phi[\tilde{f}_3^2-f_3^2-\frac{1}{r^2}\big((\tilde{f}_1^2+\tilde{f}_2^2)-(f_1^2+f_2^2)\big)] = 0 
\ee
Since $\alpha$ is a function of $r$ only, the first term and the coefficient of $\sin^2\theta \cos^2\phi$  in a second term of Eq.~\eqref{15} must individually vanish, giving us two different copy equations
\be 
&&\tilde{f}_1^2+\tilde{f}_2^2=f_1^2+f_2^2 \  \  \Rightarrow \  \ 
f_2+\frac{1}{2} = -f_1 \cot\frac{\alpha}{2} \label{16}\\
&&\tilde{f}_3^2=f_3^2 \ \  \Rightarrow  \  \ 
 f_3 \dot{\alpha} + \frac{1}{4}\dot{\alpha}^2 = 0  \ 
 \Rightarrow \hspace{0.5 cm}   \dot{\alpha}=0  \   \  \text{or} \  \   \dot{\alpha}=-4 f_3  \label{17}
\ee

For a non-trivial copy to exist, Eq.~\eqref{16} has to be satisfied, with a parameter $\alpha$ in it satisfying either of  two eq.s in  Eq.~\eqref{17}.
There are two choices to make since $f_1$ and $f_2$ are arbitrary functions. 
\begin{enumerate} 
\item\label{1} \ $f_2+\frac{1}{2} \neq -f_1\cot\frac{\alpha}{2}$
\item\label{2} \ $f_2+\frac{1}{2} = -f_1 \cot\frac{\alpha}{2}$. 
\end{enumerate} 
 If 
 \begin{equation}
  f_2+\frac{1}{2} \neq -f_1 \cot\frac{\alpha}{2}
 \end{equation}
  then  it is clear that no copy exists for this choice.\\
  However, if   
\begin{equation}\label{21}
f_2+\frac{1}{2} = -f_1 \cot\frac{\alpha}{2}
\end{equation} then we encounter  two copies  corresponding to eq.s $\dot{\alpha}=0$ and $\dot{\alpha}=-4f_3$. They are obtained by putting  Eq.~\eqref{21} in the transformation~\eqref{tra}
\be
\tilde{f_1}&=& -f_1  \nonumber\\
\tilde{f_2}&=& f_2 \nonumber 
\ee
Therefore for $\dot{\alpha}=0$  \ (putting $\dot{\alpha}=0$ back in transformation~\eqref{tra} ), we obtain
\be
\tilde{f_1}&=& -f_1  \nonumber\\
\tilde{f_2}&=& f_2 \nonumber\\
\tilde{f_3}&=& f_3   \nonumber
\ee
which yields a copy
\begin{equation}\label{1111}
\tilde{A}_j^j = A_j^j-2if_1 \Big(\frac{1}{r}-\frac{x_j^2}{r^3}\Big) 
\end{equation}
\begin{equation}\label{1112}
\tilde{A}_k^j = A_k^j+2if_1 \frac{x_jx_k}{r^3}
\end{equation}
However, on a compact manifold $\mathbb{S}^3$  this copy no longer exists. Because 
   $\dot{\alpha}=0 \Rightarrow \alpha = const.$ everywhere including infinity. Setting  $\alpha(r) = \alpha(\infty) = 4\pi n$, for which the copy Eq.~\eqref{21} implies  $f_2 = \infty$ everywhere giving a copy which is also $\infty$. 
    We want finite  copies  of $A^j_k$ which is well behaved and finite  at finite distances, which is not possible for $\dot{\alpha}=0$ on  $\mathbb{S}^3$. Therefore, the  Eq.~\eqref{21} is not valid on $\mathbb{S}^3$, thus the copy vanishes on it. The other possibility is $f_1=0$ everywhere for a given $f_2(r)$ but by Eq.s~\eqref{1111},\eqref{1112} we get the original configuration as a copy.
  
  Now, we are left with only one copy which corresponds to
\begin{equation}\label{22}
 \dot{\alpha}=-4f_3 \Rightarrow \alpha=-4\int f_3 \  dr + const.
\end{equation}
Putting $\dot{\alpha}=-4f_3$ back in transformation~\eqref{tra}, we get
\be
\tilde{f_1}&=& -f_1  \nonumber\\
\tilde{f_2}&=& f_2 \nonumber\\
\tilde{f_3}&=& -f_3   \nonumber
\ee
which yields a copy
\be
\tilde{A}_j^j = -A_j^j\\
\tilde{A}_k^j = -A_j^k
\ee
 It can also be removed on $\mathbb{S}^{3}$. We recall   boundary conditions~\eqref{23}. Since $f_3 \rightarrow 0$  faster than $\frac{1}{r} \ \ \text{as} \ \ r \rightarrow \infty$, Eq.~\eqref{22} implies that $\alpha(\infty)=const.$. As for the previous copy, we  set $ \alpha(\infty)= 4\pi n$ for which the  Eq.~\eqref{21} implies
 $ f_2 \rightarrow   \infty \ \ \text{as} \ \ r \rightarrow \infty$. Hence it is clear that on $\mathbb{S}^{3}$, Eq.~\eqref{21} is an obstruction for the boundary condition on $f_2$ (Eq.~\eqref{23}) to be satisfied therefore not valid. Therefore, this copy does not exist on $\mathbb{S}^{3}$.

  The result is  true under stronger general boundary conditions such as $\frac{1}{r^2}$, $e^{-r}$ and all cases where $\cot\frac{\alpha}{2}\rightarrow \infty$ faster than $f_1$ decays. Similarly, it can be shown that the  condition for other two components, $ \tilde{A}_{i}^2 \tilde{A}^{i 2}=A_i^2A^{i 2}$  \ and \  $\tilde{A}_{i}^3 \tilde{A}^{i 3}=A_i^3A^{i 3}$,  produce same two equations for copy. 
 
 Whereas for coulomb gauge, we have~\cite{2}
\begin{align}
\frac{\partial A_i}{\partial x_i} = \hat{n}(\dot{f}_3 + \frac{2}{r}f_3 - \dfrac{2}{r^2}f_1)
\end{align}
Because Pauli matrices $\sigma_a$ are  unit vectors in $2 \times 2$ matrix space,  the condition
\begin{align}
\frac{\partial \tilde{A}_i^a}{\partial x_i} = \frac{\partial A_i^a}{\partial x_i}
\end{align}
for all three components yields the equation 
\begin{align}
\ddot{\alpha} + \frac{2}{r}\dot{\alpha}-\frac{4}{r^2}\Big( (f_2+\frac{1}{2})\sin \alpha + f_1 \cos \alpha\Big) = 0
\end{align}
This equation is known to be solvable and therefore the ambiguity exists even on $\mathbb{S}^{3}$.

\section{\tmem{BRST} symmetry in quadratic gauge  }\label{s3}
    In this section, we prove that this theory is \tmem{BRST} invariant.   We begin by writing {\tmem{BRST}} transformations in the quadratic gauge:
\begin{subequations}\label{eq: 4 }
\begin{align}
&\delta c^d   = \frac{\omega}{2} f^{d b c} c^b c^c  \\
&\delta\overline{c^d}  =  \frac{2 \omega}{g } F^d\\
&\delta A_{\mu}^d  = \frac{\omega}{g} ( D_{\mu} c)^d\\
&\delta F^d = 0
\end{align}
\end{subequations}
Nilpotency of the transformations~\eqref{eq: 4 } can be easily checked.
Under these transformations, variation of the $\mathcal{L}_{\tmop{eff}}$ in  Eq.~\eqref{eq: 9 } is as follows
\be \label{eq. 8}
\delta \mathcal{L}_{\tmop{eff}} = &\hspace{.5 em}& \delta \left(
\frac{\zeta}{2}F^{a 2} + F^a\ A^a_{\mu} A^{\mu a}
 - \overline{c^a}A^{\mu a} (D_{\mu} c)^a \right)  \hspace{2 cm} \Big(\delta \mathcal{L}_{\tmop{YM}} = 0\Big)   \nonumber \\
 =& &\frac{2 \omega}{g } F^a A^{\mu a} ( D_{\mu} c)^a - 
 \frac{2 \omega}{g } F^a A^{\mu a} ( D_{\mu} c)^a \noplus   \nonumber \\
&-& \frac{\omega}{g} 
\overline{c^a} ( D_{\mu} c)^a ( D^{\mu} c)^a \hspace{2 cm}\Big(  \text{We have used} \ \delta (D_{\mu} c)^a = 0 \Big) \nonumber\\
 =&-& \frac{\omega}{g}  \overline{c^a}  (D_{\mu} c)^a ( D^{\mu} c)^a\\
=  & \hspace{0.5em} 0 &  \hspace{2 cm}\Big((D_{\mu} c)^a \ \ \text{is a grassmann variable}\Big)\nonumber
\ee
Thus, we prove  that the theory is \tmem{BRST} invariant.

\section{conclusion}
We discussed a particular quadratic gauge, which is a Lorentz invariant  algebraic gauge. We worked out an example of spherically symmetric  configuration in the quadratic gauge and  proved that the configuration with a proper boundary condition does not have any copy on $\mathbb{S}^3$. Thus, we provided one  example where an algebraic gauge is compatible with the boundary condition on the fields and the compactification of the space is possible in an algebraic gauge. We also proved that the  theory is \tmem{BRST} invariant.

\section*{Acknowledgment}
Haresh is sincerely thankful to Professor Urjit A. Yajnik for useful comments on the subject.

\end{document}